\documentclass[preprint2,numberedappendix]{emulateapj-rtx4}
\usepackage{amssymb,amsmath}
\usepackage{bm,graphics,epsfig}
\usepackage{color}
\usepackage{subfigure}
\usepackage{ulem}
\graphicspath{{./fig/}{./png/}}
\usepackage{url}
\def \del2z {\partial^{2}_{z}}

\def \uu {{\bm U}}
\def \ap {R_{\rm  SB}}
\def \a {R}

\def \vp {{\bm v}_{\rm SB}}

\newcommand{\eq}[1]{(\ref{#1})}

\newcommand{\Eq}[1]{Equation~(\ref{#1})}

\newcommand{\bra}[1]{\langle #1\rangle}
\newcommand{\ddt}[1]{\frac{d #1}{dt}}
{}

\def \Rey  {\mbox{Re}}
\def \Rep  {\mbox{Re}_{ \rm SB}}

\def \cs  {c_{\rm s}}

\def \taup {\tau_{\rm SB}}

\def \tauO {\tau_{\rm orb}}
\def \RO {R_{\rm orb}}

\def \Ok  {\Omega_{\rm kepler}}
\def \uk  {v_{\rm kepler}}
\def \Stk {\mbox{St}_{\rm kepler}}

\def \mach {\mathcal{M}}
\def \vwind {v_{\rm wind}}

\def \tauL {\tau_{\rm L}}
\def \rhop {\rho_{\rm SB}}
\def \rhog {\rho_{\rm g}}
\def \vn {v_{\rm n}}
\def \cs {c_{\rm s}}

\def \vzero {v_{\rm 0}}

\def \St   {\mbox{St}}
 
\def \Np {N_{\rm p}}
\newcommand{\Fig}[1]{Fig.~\ref{#1}}

\def\url#1{\textcolor{blue}{\protect\small\sf #1}}

\newcommand{\yjfm}{J. Fluid Mech.}
\newcommand{\icarus}{Icarus}
\def\mnras{MNRAS}

\begin{document}
%\title[Planetesimals by collisional fusion]{Can planetesimals form by collisional fusion?}
\title{Can planetesimals form by collisional fusion?}
\author{Dhrubaditya Mitra$^1$, 
       J.S. Wettlaufer$^{1,2}$, and  
        Axel Brandenburg$^{1,3}$
}
\affil{
$^1$Nordita, KTH Royal Institute of Technology and Stockholm University,
Roslagstullsbacken 23, SE-10691 Stockholm, Sweden\\
$^2$Yale University, New Haven, CT, USA\\
$^3$Department of Astronomy, AlbaNova University Center,
Stockholm University, SE-10691 Stockholm, Sweden\\
}
%%%%%%%%%%%%%%%%%%%%%%%%%%%%%%%%%%%%%%%%%%%%%%%%
%\date{\today}
\date{\today, $ $Revision: 1.61 $ $}
%\pagerange{\pageref{firstpage}--\pageref{lastpage}} \pubyear{}

%\maketitle

%\label{firstpage}

\begin{abstract}

As a test bed for the growth of protoplanetary bodies in a turbulent
circumstellar disk we examine the fate of a boulder using direct numerical
simulations of particle seeded gas flowing around it.
We provide an accurate description of the flow by imposing no-slip and
non-penetrating boundary conditions on the boulder surface using the
immersed boundary method pioneered by Peskin~(2002).
Advected by the turbulent disk flow, the dust grains collide with the
boulder and we compute the probability density function (PDF) of the
normal component of the collisional velocity.
Through this examination of the statistics of collisional velocities we test the recently
developed concept of collisional fusion which provides a physical basis
for a range of collisional velocities exhibiting perfect sticking.
A boulder can then grow sufficiently rapidly to settle into a Keplerian
orbit on disk evolution time scales.
\end{abstract}

%\begin{keywords}
%planet formation
%\end{keywords}
\section{Introduction}

\subsection{Accretion Disks and Protoplanets}
 
Planet formation is hypothesized to occur through the growth of
protoplanetary bodies formed from gas, dust and ice grains in an accretion
disk around a central star \citep{Arm10}. The complex scenario of the
planet formation process involves the following four stages. 
Firstly, the initial collapse of interstellar gas to create the central 
protostar ($\sim0.1$\,My);
secondly, the slow accretion of mass onto the star and the formation of
primary planetesimals within the evolving accretion disk ($\sim$ My);
thirdly, a phase ($\sim$ My) of reduced accretion rate allowing the
photoevaporative wind to divide the disk into an inner and an outer region
at a radius determined by the ratio of the stellar accretion rate
to the mass loss rate due to photoevaporation;
finally, there is a \textit{clearing phase} ($\sim0.1$\,My) during which
the inner disk accretes onto the star while the lightest elements of
the outer disk are removed due to direct exposure to photoevaporative
UV flux.
Recent cosmochemical evidence reveals that the long held view of a
$\sim$ My age difference between Ca-Al-rich inclusions (CAIs) within
carbonaceous chondrite meteorites and chondrules within chondrites can be
refuted \citep{Connelly12}.
To the extent that these data demonstrate commensurability over disk
lifetime scales of CAI and chondrule formation, the detailed transient
development of matter within circumstellar disks becomes all the more
compelling for studies that can isolate essential physical processes.
Here we focus on fundamental aspects of the second stage above.  This stage
is crucial for understanding how the material that forms the building
blocks of planets can organize into bodies that thwart the radiative
pressure effects in the subsequent stages that sweep the disk of small
particles and gas.\footnote{A different hypothesis originally due to
Safronov and Goldreich and Ward \citep[see e.g.][for a review]{gol+lit+sar04, Arm10,
Youdin10} leads to planetesimals by the gravitational collapse of the
disk material.  We do not consider this here. 
}

The accretion disk is treated as a two phase system defined by a fluid
phase (`gas') and solid particles (`dust') advected by the fluid.
Ubiquitous attractive long range van der Waals and electrostatic 
interactions facilitate the agglomeration and growth of small (micron
or smaller) dust grains that are brought into proximity by the turbulent
flow of the gas.
However, depending on the material and the mechanical and thermodynamic
conditions of a particle-particle collision, sticking (through a number
of mechanisms), fragmentation, or bouncing will determine the fate and
the size distribution of accreting matter from the small scales upward
\citep{blu+wur08, wet10, Zsom11}.

Because the central star creates a radially decaying pressure gradient,
the gas moves at a slightly sub-Keplerian speed.
Thus,  depending on the position in the disk, there are a range of
particle sizes that experience a strong ``headwind'' and so lose angular
momentum, thereby driving them into the central star on time scales as
rapidly as a century \citep{Arm10, Youdin10}.
We are concerned with the long standing problem of how, when objects
grow and begin to experience the local headwind, they can accumulate
matter sufficiently quickly to slow their drift inward.
To focus the question, we examine in some detail how a meter sized object
grows by accretion of small particles mediated by turbulent flows of
the gas.

\subsection{Hydrodynamic Preliminaries}

The typical value of the ``disk Mach number'' $\mach_d$ is based on the Keplerian velocity $\uk$, which in the thin disk approximation is 
\begin{equation}
\mach_d = \frac{\uk}{\cs} \approx \frac{r}{h}, 
\end{equation}
where $r$ is the radial position in the disk and $h$ is its vertical scale height.  
At $1$\,AU, $h/r \approx 0.02$ and hence  $\mach_d \approx 50$ \citep[see, e.g.,][p.\ 40]{Arm10}. 
Now, as noted above, because the central star creates a radially decreasing
pressure gradient, the gas moves at a sub-Keplerian speed
$\vwind = \eta \uk$ where $\eta$ can be as small as $10^{-3}$ depending
on the position in the disk.

To understand the effects of the interaction between the dust and the gas,
we begin by considering a solid body of spherical shape with radius $\ap$,
moving through a gas with kinematic viscosity $\nu$ and speed
$\vwind$.
We estimate its Reynolds number as,
\begin{equation}
\Rep=\frac{\vwind \ap}{\nu} = \frac{\vwind}{\cs} \, \frac{\ap}{\lambda} \, \frac{\lambda \cs}{\nu}
  \approx \mach \, \frac{\ap}{\lambda}, 
\label{eq:Rey}
\end{equation}
where $\mach \equiv \vwind/\cs$ is the Mach number of the headwind
and $\lambda$ the mean-free-path of the gas molecules.
Importantly, for this estimate we have used the well-known expression for the
viscosity of gases $\nu \sim \cs\lambda$ \citep[see e.g.,][section 8]{Lif+Pit81}.
Now, because $\mach = \eta \mach_d$, we can have $\mach \approx 0.05$,   
and hence, so long as $\ap < \mach^{-1}\lambda\approx 20\, \lambda$,
the local Reynolds number of the solid body is less than unity.
For $\ap \sim \lambda$ the size of the solid body is well below the smallest
hydrodynamic length scale in the gas and its motion is then described
by the simple drag law
\begin{equation}
\ddt{\vp} = \frac{1}{\taup}\left( \vp - \uu \right),
\label{eq:drag}
\end{equation}
where $\vp$ is the velocity of the particle, $\uu$ is the local velocity
of the gas, and $\taup$ is the so-called stopping time describing
the deceleration of particle motion relative to the gas.
When a particle is smaller than the typical hydrodynamic length scale
in the problem, $\taup$ is given by the Epstein drag law,
\begin{equation}
\taup^{\rm Ep} = \frac{\rhop}{\rhog}\frac{\ap}{\cs}, 
\label{eq:epstein}
\end{equation}
where $\rhop$ is the material density of the solid particles and $\rhog$ is the gas density.  
When $\mach^{-1}\lambda > \ap > \lambda$, the relevant drag law is
that of Stokes and $\taup$ is given by
\begin{equation}
\taup^{\rm St} = \frac{2}{9} \frac{\rhop}{\rhog}\frac{\ap^2}{\nu} .
\label{eq:stokes}
\end{equation}
Despite the fact that when $\ap > \mach^{-1}\lambda$, the simple
drag law \eq{eq:drag} no longer describes the motion of the dust
particles, most numerical approaches to these problems
\cite[see, e.g.,][]{joh+ois+mac+kla+hen+you07,Arm10,nel+gre10,
car+cuz+hog10,car+bai+cuz11} continue to use it because a more
accurate description is computationally prohibitive.
Here we will call bodies of approximately this size ``boulders''.
The mean-free-path $\lambda$ in an accretion disk varies with radius;
e.g., according to the minimum mass solar nebula model 
$\lambda$ ranges from  $\approx 10 {\rm cm}$ at approximately $1.5$ AU 
to $\approx 10 {\rm m}$ at $10$ AU. 
Hence $\a_{\rm boulder}$ ranges from  $\sim 2\,{\rm m}$ in inner disk
regions to $\sim 200\,{\rm m}$ at about $10$\,AU. 
A more accurate approximation of the motion of such particles is given
by the Maxey-Riley equation \citep{max+ril83}, which assumes a spherical
geometry.
While the Maxey-Riley approach is appealing on fundamental grounds,
it has yet to be used in simulations of fully developed turbulence.

\subsection{Bouncing, Sticking, Fusing\label{sec:stick}}

A crucial and often-used assumption is that all collisions have a sticking
probability of unity.
Indeed, under such an assumption planetesimal growth under a wide
range of disk conditions is sufficiently rapid that there is no loss
to the central star.
Clearly, however, the probability of sticking depends, among other things,
on the collisional velocity, the material properties of the colliding
bodies, the ambient temperature, and the relative particle size.
It is a commonly accepted picture that for collisional velocities $V_c$
above a certain threshold value, $V_{\rm th} \sim 0.1$--10\,cm\,s$^{-1}$,
particle agglomeration is not possible; and elastic rebound overcomes
attractive surface and intermolecular forces \cite[e.g.,][]{cho+tie+hol93}.
However, for bodies covered with ice, experimental \citep{blu+wur08}
and theoretical \citep{wet10} studies of collisions between dust grains
and meter-sized objects have elucidated the range of collisional
velocities (which depends on the relative particle size) over which
perfect sticking occurs.
This latter work considers the basic role of the phase behavior
of matter (phase diagrams, amorphs and polymorphs) in leading to
so-called \textit{collisional fusion}.
In this fusion process, a physical basis for efficient sticking is
provided through collisional melting/amphorphization/polymorphization
and subsequent fusion/annealing to extend the collisional velocity range
of sticking to $\Delta V_c \sim$ 1--100\,m\,s$^{-1} \gg V_{\rm th}$, which
encompasses both typical turbulent rms (root-mean-square) speeds and the velocity differences
between boulders and small grains $\sim1$--50\,m\,s$^{-1}$.
Moreover, bodies of high melting temperature and multicomponent materials, such as silicon and olivine, can fuse
in this manner depending on the details of their phase diagrams.
Hence, in principle, the approach provides a framework for sticking from
the inner to the outer nebula.  Here, we explore the influence of such
a range, $\Delta V_c$, on the growth of a boulder in a simulated disk.

\subsection{Summary of Approach}

The fate of the boulder is studied from a reference frame fixed to it, while the gas flows around it. 
We provide an accurate description of the flow by performing a direct numerical simulation (DNS) 
with no-slip and non-penetrating boundary conditions on the boulder surface using a numerical technique 
called \textit{Immersed Boundary Method} \citep{Peskin02}. 
Hence, there is no ad hoc approximation involved in describing the mutual
interaction between the boulder and the gas flow.
However, at present, it is computationally prohibitive to solve for more than one boulder using this DNS scheme.
Consequently we focus our study on the flow mediated collisions between
one boulder and many ``effectively'' point sized dust grains whose sizes are much smaller than $\a_{\rm boulder}$. 
Our principal approximations in treating the motion of the dust grains are 
(a) to use \Eq{eq:drag} and 
(b) to ignore the back-reaction of the dust grains onto the flow.  
Advected by the turbulent disk flow, the dust grains collide with the boulder and we compute the PDF of the 
normal component of the collisional velocity. 

\section{Model}

The mechanism of formation of planetesimals from dust grains is modeled by the same tools that are used to 
study, for example, hydrometeor growth in the terrestrial atmosphere,  namely the coagulation/fragmentation 
equations of \cite{smo16}; see, e.g., \cite{Arm10}, for a recent review. 
The Smoluchowski equations are integro-differential equations that require two crucial ingredients:
the probability distribution function of relative collisional velocities of the bodies in question and 
their sticking efficiency. 
The former, particularly for the inner disk region, is strongly influenced by turbulence. 
Recently, there has been significant progress in calculating
the statistical properties of individual particle velocities
\citep{car+bai+cuz11,nel+gre10} and, perhaps more importantly, pairwise
relative velocities \citep{car+cuz+hog10} from direct numerical
simulations.
Similar results have also been obtained from both phenomenological
\citep{orm+cuz07,cuz+hog03} and shell \citep{hub12} models of turbulence.
While these approaches provide key insights and intuition, they also leave
open aspects with which the strategy we take is not burdened, such as
(a) the use of the simple drag law~\eq{eq:drag} to describe the motion of boulders,
(b) the ability to obtain only the root-mean-square collision velocity, rather than the PDF of collision velocities \cite[][are exceptions]{car+cuz+hog10,hub12},
(c) not modeling actual collisions, so that collisional velocities are inferred from looking at relative velocities
at small distances. 
To calculate the PDF of collisional velocities between a boulder and small dust grains, such approximations
may be too simplistic because of the presence of a boundary layer around the larger object. 
Indeed, \cite{gar+mer+gal+olc13} have recently pointed out the importance of using the PDF of collisional velocities instead of simply the root-mean-square value.  However, taking this into account in a global (or even local) simulation of a disk is computationally prohibitive.   Therefore, we take an initial modest step to try and understand such collisions by solving the equations of motion for weakly compressible fluids in two dimensions
with a circular object--the boulder--inside.
We ignore two classes of collisions, (a) between dust grains themselves, and 
(b) between two or more boulders. 

\subsection{Numerical method}
%----------
\begin{figure}
\includegraphics[height=0.9\columnwidth,angle=90]{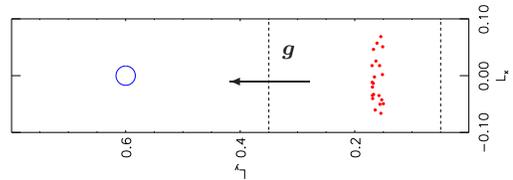} 
\put(-80,50){\vector(-1,0){30}}\thicklines
\put(-90,60){$ {\bm g}$}
\caption{A sketch of our computational domain. The domain is divided into two halves. 
The left half contains the ``boulder'' sketched by the blue circle.  In the right half the fluid is 
acted upon by an external white-in-time force which is non-zero only in the part of the domain 
limited by the two dashed lines.  The turbulence thus generated is moved toward the ``boulder'' by the
action of weak body force ${\bm g}$ along the arrow shown in the figure. The body force does not act directly
on the particles, which are introduced continuously in a small area in the right half of the domain. 
Initial positions of a few particles are shown as red dots.}
\label{fig:domain}
\end{figure}
%-------------
Our computational domain is a rectangular box divided into two equal parts (\Fig{fig:domain}).
In the right half, fluid turbulence is generated by external forcing that is non-zero between the two
dashed lines shown in \Fig{fig:domain}. 
The turbulence thus generated is moved toward the ``boulder'' by the
action of a body force ${\bm g}$ in the direction of the arrow shown in the figure.
This body force is responsible for generating a mean flow, which models the head-wind
faced by a boulder--the circular object at the left half of the domain. 
The boundary layer around the boulder is fully resolved by imposing non-penetrating and no-slip boundary 
conditions using the immersed boundary method. 
After the flow has reached a stationary state, we introduce $\Np = 2 \times 10^{4}$ particles into the 
right half of the domain as depicted in \Fig{fig:domain}.
The motion of these particles obeys the simple drag law,
\begin{equation}
\ddt{{\bm v}_{\rm p}} = \frac{1}{\tau_{\rm p}}\left({\bm v}_{\rm p}  - \uu \right),
\label{eq:dragp}
\end{equation}
with the characteristic drag time of the ``dust particles'' $\tau_{\rm p}$.
As noted before and as is clear from context, no such assumption need be made for the boulder. 
The back-reaction from the dust grains to the gas is ignored. 
When a dust grain collides with the boulder it is removed from the simulation and a new dust grain is introduced in the
right half of the domain.
We use the {\sc Pencil Code}%
\footnote{\url{http://pencil-code.googlecode.com/}} 
in which the immersed boundary method was first implemented by \cite{hau+kra10}. 

\subsection{Parameters}

The characteristic large-scale velocity is the root-mean-square velocity in the streamwise direction, 
$\vwind\equiv\bra{v^2_y}^{1/2}$. 
We always use the Reynolds number corresponding to the central solid body, defined by 
\begin{equation}
\Rep \equiv \vwind\ap/\nu.
\label{eq:ReSB}
\end{equation}
And the  Stokes number of the ``dust particles'' is defined by
\begin{equation} 
\St \equiv \tau_{\rm p}/\tauL,
\label{eq:St}
\end{equation}
where $\tauL = L_y/\vwind$ with $L_y$ being the length of our domain along the streamwise direction; from right to left in \Fig{fig:domain}.
By virtue of limiting our simulations to two dimensions we can access a larger range of particle Reynolds numbers $\Rep$,
from $30$ to $1000$ with resolutions ranging from ${\tt 128}\times{\tt 512}$ to ${\tt 512}\times{\tt 2048}$ grid points. The surface of the boulder is resolved with ${\tt 100}$ to ${\tt 400}$ grid points. 
\section{Results}
\label{sec:results}
%----------
\begin{figure}[t]
\includegraphics[width=0.9\columnwidth]{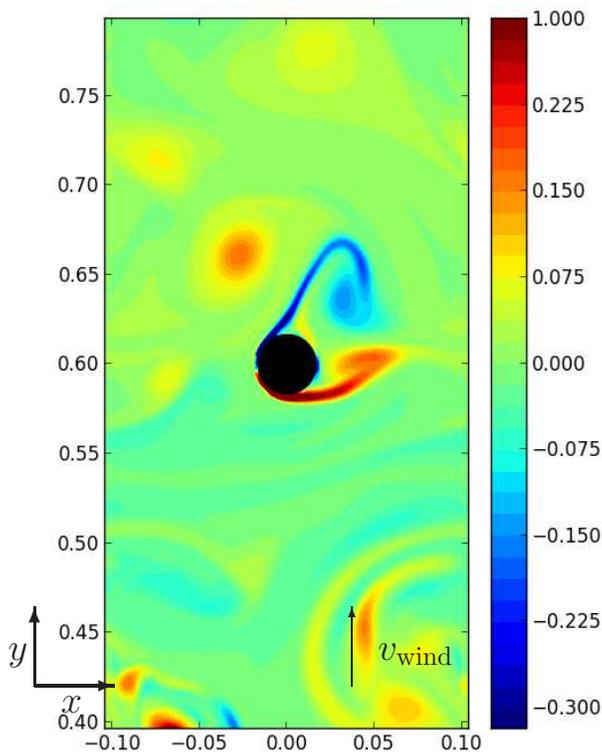} 
\put(-100,50){\vector(0,1){30}}\thicklines
\put(-90,60){\Large{$\vwind$}}
\put(-220,50){\vector(0,1){30}}\thicklines
\put(-230,60){\Large{$y$}}
\put(-220,50){\vector(1,0){30}}\thicklines
\put(-210,40){\Large{$x$}}
\caption{Contour plot of vorticity in the upper half of our domain. The black circle at the center of the domain is the circular object.
The arrow shows the time and space averaged direction of $\vwind$. }
\label{fig:vort}
\end{figure}
%-------------
%----------
\begin{figure}[t]
\includegraphics[angle=90,origin=c,width=0.95\columnwidth]{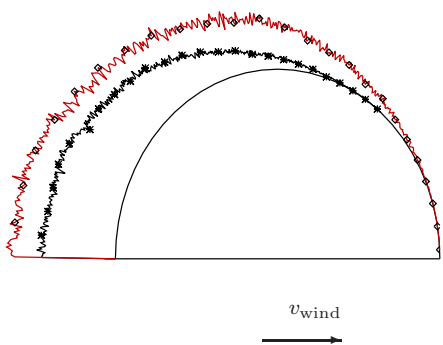} 
\put(-100,50){\vector(1,0){30}}\thicklines
\put(-90,60){$\vwind$}
\caption{Plot showing how the boulder would grow if all collisions were perfectly sticky. The arrow shows the direction of $\vwind$. The growth for two
different runs (a) $\Rep\approx 29$, $\St\approx 0.5$ ($\ast$), and (b) $\Rep\approx 1000$, $\St\approx 0.6$ ($\square$),
for the same total time duration are shown. 
The inner semi-circle shows the initial surface of the boulder. 
}
\label{fig:pdf_angle}
\end{figure}
%-------------
A representative snapshot of the vorticity field is given in \Fig{fig:vort}. 
A movie, available online at: \url{http://www.youtube.com/watch?v=-Fr5Q2Kp0wo}, shows that, although over a spatiotemporal average there is a streamwise mean flow on the boulder, there are large fluctuations.
At a particular instant the direction of the gas velocity at the boulder surface can deviate significantly from the streamwise direction. 
Furthermore we observe that most of the collisions do not occur at the front face of the boulder but there is a
significant number of collisions that deviate from centrality; see \Fig{fig:pdf_angle}.  
Note, however, that there are almost no collisions on the backside of the 
boulder. 
A clear implication of this is that, for perfect sticking of all collisions, an initially spherical boulder evolves into a 
non spheroidal body and hence may begin to tumble in the disk. 

%\eject

\subsection{PDF of collisional velocities}
%----------
\begin{figure}
\includegraphics[width=0.90\columnwidth]{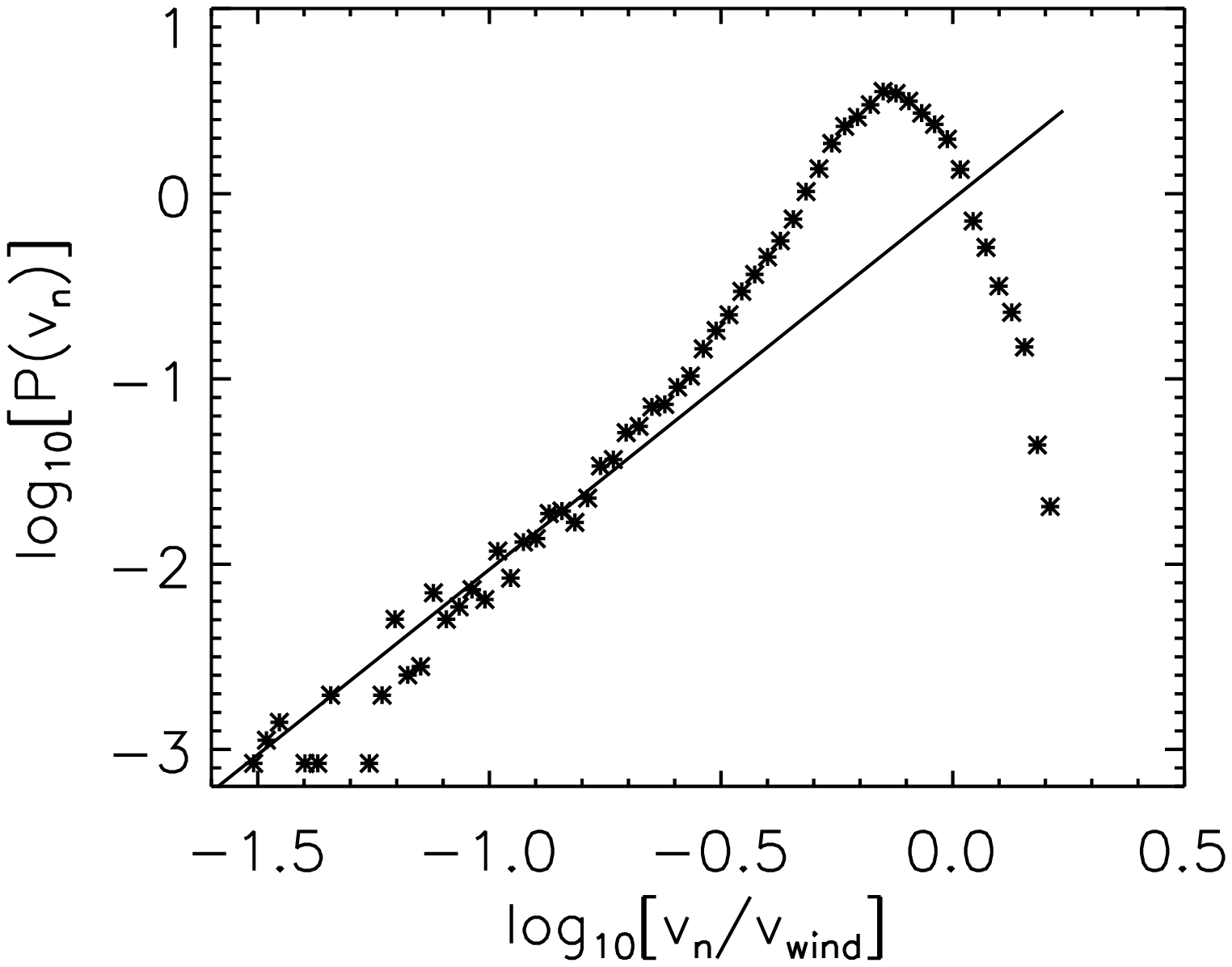}  
\put(-150,100){\Large{(a)}}\\
\includegraphics[width=0.90\columnwidth]{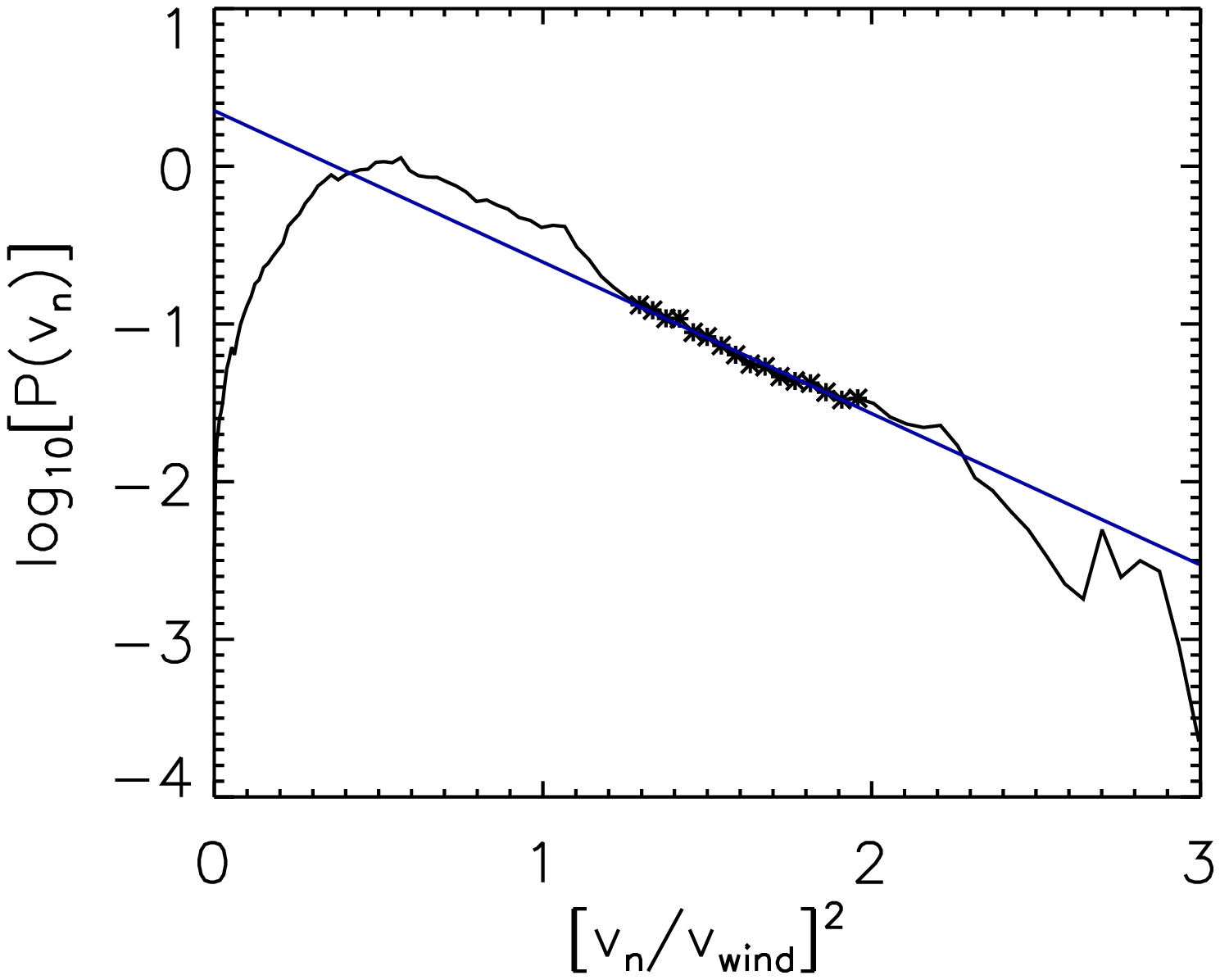} 
\put(-150,60){\Large{(b)}}\\
\includegraphics[width=0.90\columnwidth]{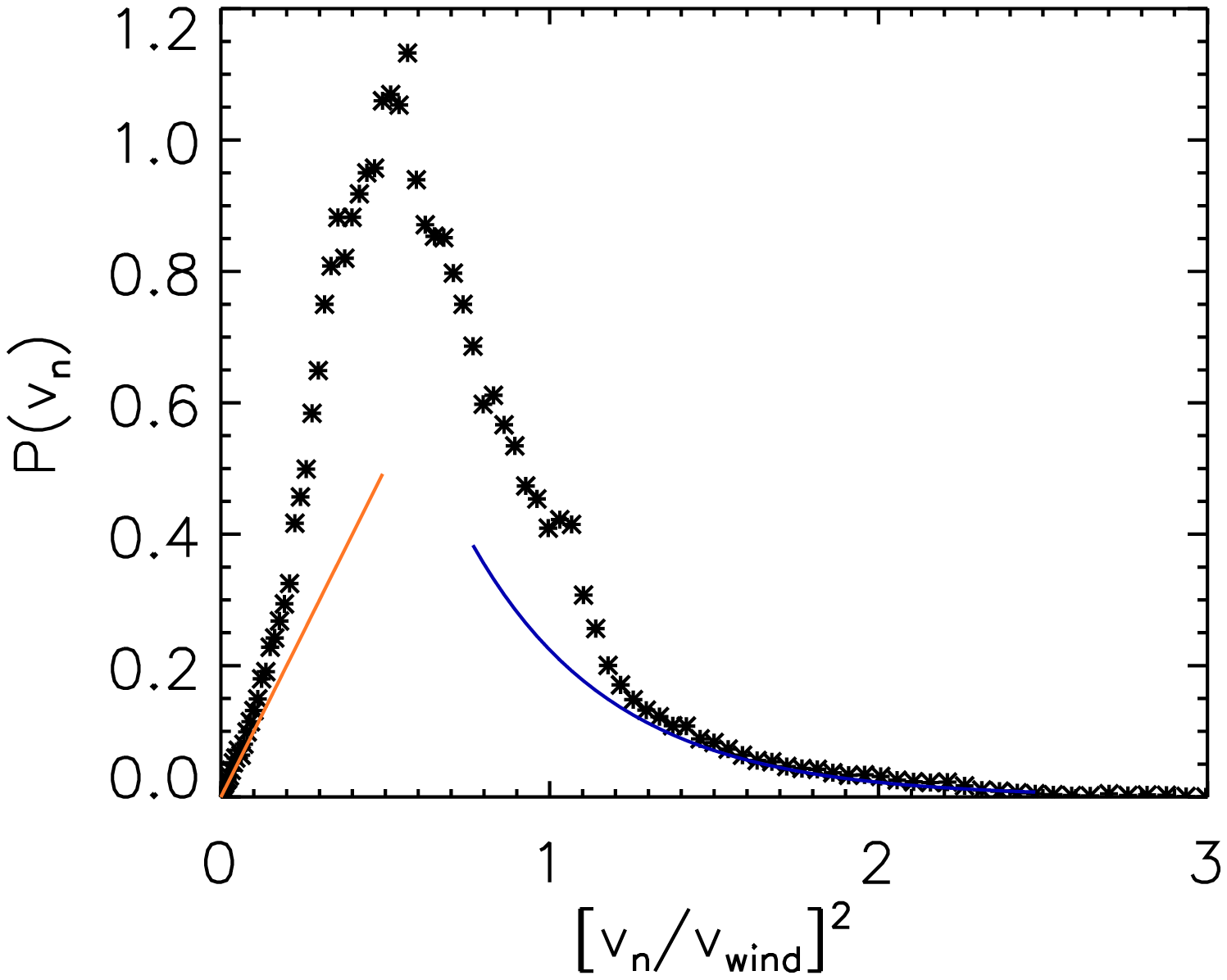} 
\put(-80,80){\Large{(c)}}
\caption{ PDF of collisional velocities for $\Rep\approx 1000$,$\St\approx 0.6$.
(a) Log-log (base $10$) plot for  small $\vn$; $P(\vn) \sim (\vn/\vwind)^2$, the straight line has a slope of $2$. 
(b) Semi-log (base $10$) at large $\vn$. The straight line, which is a fit to the points denoted by the symbol $\ast$, has slope $0.96$ 
(c) The PDF with the two approximations at small and larger $\vn$ plotted together. 
}
\label{fig:pdf_Remax}
\end{figure}
%----------------
%----------
\begin{figure}
\includegraphics[width=0.90\columnwidth]{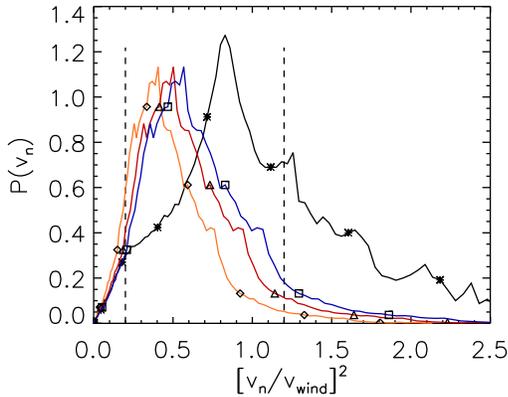} 
\caption{Probability distribution function, $P(v_n)$, of the normal component of collisional velocity versus $(v_n/\vwind)^2$ for four different
runs: (a) $\Rep\approx 29$, $\St\approx 0.5$ ($\ast$),
(b) $\Rep\approx 69$, $\St\approx 1$ ($\square$),
(c) $\Rep\approx 516$,$\St\approx 0.7$ ($\vartriangle$),
(d) $\Rep\approx 1000$,$\St\approx 0.6$ ($\lozenge$).
}
\label{fig:pdf_Redep}
\end{figure}
%----------------
%----------
\begin{figure}
\includegraphics[width=0.9\columnwidth]{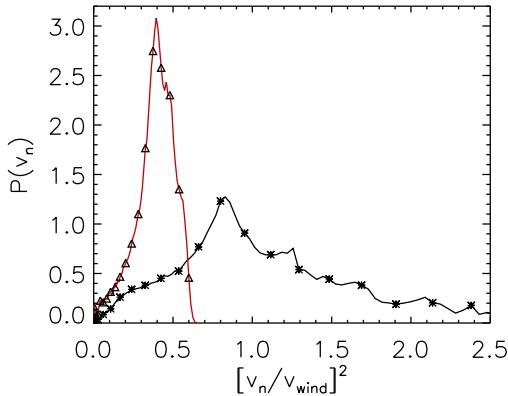} 
\caption{Probability distribution function, $P(v_n)$, of normal component of collisional velocity versus $[v_n/\vwind]^2$ for two different
Stokes numbers; $\St\approx 0.5$ ($\ast$), and  $\St\approx 0.1$ ($\vartriangle$), for $\Rep\approx 29$. 
}
\label{fig:pdf_st}
\end{figure}
%-------------
The criterion for a collision is that the distance between a dust grain and the boulder becomes less than a grid point.  After this collision we remove the dust grain
from the simulation.  For $\Rep\approx 1000$ in \Fig{fig:pdf_Remax}; we plot the PDF, $P(\vn)$, of the component 
of the velocity of the dust grain normal to the surface of the boulder, $\vn$.
At small $\vn$,  $P(\vn) \sim \vn^2 $ (\Fig{fig:pdf_Remax}a) and at large $\vn$ the 
fall off is $\sim\exp[-(\vn/\vzero)^2]$ (\Fig{fig:pdf_Remax}b). 
However, as shown in \Fig{fig:pdf_Remax}(c), over the whole range it is difficult to fit the PDF with a Maxwellian 
distribution. 

Now we consider how the PDF changes as the Stokes and Reynolds numbers of the flow change.
We vary the Reynolds number by changing the viscosity of the flow.
Hence, a change in Reynolds number also changes $\vwind$, and this leads to a change in the 
Stokes number\footnote{As we change viscosity holding all other variables, including the 
body force ${\bm g}$,  constant, $v_{\rm wind}$ also changes. This changes $\tauL = L_y/v_{\rm wind}$
which consequently changes $\St$ through \eq{eq:St}. }.  
Therefore, in our approach to the numerical treatment of the flow, it is not possible to perform a systematic study of the Reynolds number dependence of the PDF at fixed Stokes
number.  However, in order to produce an effective treatment of such a circumstance, we present in \Fig{fig:pdf_Redep} the PDFs for different Reynolds numbers wherein the Stokes numbers 
are not too different from each other. 
We see that for small $\Rep$ the peak of the PDF lies very close to $\vwind$, but as $\Rep$ increases the peak moves to smaller velocities by only a very small amount. Although $\Rep$ changes by almost a factor of
$20$ the position of the peak (normalized by $\vwind$) only changes from $0.6$ to $0.3$. 
A more dramatic change is observed in the PDF as the Stokes number is changed from $0.5$ to $0.1$ when $\Rey\approx 29$ is held fixed, as shown in \Fig{fig:pdf_st}. 
In particular, the tail of the PDF at high $\vn$ is severely cut off as the Stokes number is decreased by a factor of $5$. 
One can understand this as follows;  when the Stokes number decreases, the dust grains begin to follow streamlines and hence never collide with the boulder.  The implication of this is clearly that  
a \textit{smaller Stokes number} implies  a \textit{smaller number of high-impact collisions}.  
Nevertheless, the most striking result for the problem at hand is the insensitivity of  the PDF to $\Rep$ and $\St$. 

\subsection{From DNS to disk astrophysics}
Our simulations take place in the reference frame of the
boulder. Although the boulder is also comoving with the local gas with 
velocity $\uk$, the head wind corresponds to $\vwind$ in our simulations, thereby setting the velocity scale. 
The radius of the boulder  is taken to be $\approx 10\, {\rm m}$. 
The magnitude of the headwind in the disk is estimated to be 
$\vwind  \approx 10^{-3} \uk \approx 3 \times 10^3\,{\rm cm\,s}^{-1}$ \cite[see e.g.][page 130]{Arm10}. 
In the astrophysical literature it is common to non-dimensionalize $\tau_{\rm p}$
with $\Ok$, the Keplerian frequency, to define the orbital Stokes number, $\Stk$.
Here, we use the largest eddy time scale $\tauL = L_y/\vwind$, to obtain $\St$.
These two Stokes numbers are related by 
\begin{equation}
\Stk = \St\frac{\tau_L}{\tauO} 
\end{equation} 
where $\tauO$ is the characteristic time scale of the Keplerian orbit is defined by
\begin{equation}
\tauO = \frac{\RO}{\uk}
\end{equation}
where $\RO$ is the orbital radius. Using the definition of the two time scales $\tauL$ and 
$\tauO$, we obtain the ratio of the two Stokes numbers to be
\begin{equation}
\frac{\Stk}{\St} = \frac{L_y}{\RO} \frac{\uk}{\vwind} \approx 10^{-5} 
\end{equation}
where we have used $\RO=1 {\rm AU}$, $L_y = 50\ap \approx 500 m$, and
$\vwind = \eta \uk$ with $\eta=10^{-3}$. 
We have used Stokes number ranging from $\St = 0.1$ to $2$ which in turn
gives $\Stk \approx 10^{-6}$ to $2\times 10^{-5}$. 
We use the same conventions used in the Supplementary Information of \cite{joh+ois+mac+kla+hen+you07} to convert 
the value of $\Stk$ to a radius of the dust grain; this implies that our ``dust particles'' are of the size of tenth of millimeters
or smaller.  Clearly the ``dust particles'' are smaller than hydrodynamic scales, and hence it is justified to 
consider them as point objects whose motion are described by the Epstein drag law. 

\subsection{Collisional fusion}

The PDFs of collisional velocities show that, irrespective of the Reynolds number and the Stokes number within the range considered by us, most collisions occur at velocities rather near to $\vwind$.
To illustrate this in \Fig{fig:pdf_Redep} we have drawn two vertical dashed lines at 
$(\vn/\vwind)^2=0.2$ and 
$(\vn/\vwind)^2=1.2$.
The area under the PDF between the two lines includes approximately $95\%$ of the total number of collisions. 
Translated to parameters in the disk, this implies that, if there is a mechanism by which dust grains with velocities
ranging from $0.2\,\vwind$ to $1.2\,\vwind$ would stick to a boulder, then we could consider $95\%$ of collisions to
have a perfect sticking probability. 

Roughly speaking, this implies a range of velocities $6-36\,{\rm m\,s}^{-1}$.
These collisional velocities are far too high for the bodies to fuse by
attractive intermolecular forces.
An alternative scenario by which the colliding bodies can fuse at high
speed has been suggested by \cite{wet10}.
As discussed in section \ref{sec:stick}, the very high local pressures
that occur during a collision can lead to phase change.
If, when the pressure begins to relax during rebound the momentarily liquified (or
disordered) interfacial material re-freezes (or anneals) before particle
separation, then fusion can occur.
The idea was demonstrated when the colliding bodies are covered by ice,
but the theory is generally applicable to all materials whose phase
diagram is known in detail. An example of the process in a high melting 
temperature material (silicon) was noted in \cite{wet10}. 
Hence, whether the range of collisional velocities over which such
process can occur in a material such as olivine matches with the range we
find here is a topic of ongoing research.
Note that here the particle Reynolds number $\Rep$ varies linearly with
the particle radius but the range over which most of the collisions occur
does not depend sensitively on $\Rep$, and hence not on the particle
radius.
Thus, runaway growth of the boulder through the accretion of dust
grains is a viable mechanism in areas of the disk where collisional
fusion can operate in the range we obtain.

\section{Discussion and Conclusion}
To describe the motion of micron sized dust grains in a protoplanetary
disk the simple drag law of \Eq{eq:dragp} is sufficient.
Theoretical estimates \cite[see e.g.,][p.\ 120]{Arm10} suggests that micron sized dust particles in the inner 
disk (about $5$\,AU) can grow up to a size of $10$'s of centimeters if we assume that the presence of turbulence increases 
the number of collisions and that almost all collisions result in coagulation by long-ranged intermolecular forces.  
But the process that allows them to continue to grow to the size of planetesimals is not well understood.
As the dust grains grow, at some stage they become boulders and their local Reynolds number exceeds unity.
At this stage we need a more accurate description of their interaction with the gas than the one provided by
\eq{eq:drag}. 
Here we provide such a description of a boulder colliding with dust grains by using the immersed boundary method of \cite{Peskin02}. 
Remarkably, we find that the PDF of collisional velocities depends weakly on $\Rep$ and $\St$.  In particular,
we find that, if collisional fusion between dust grains is possible in the range of collisional velocities $\Delta V_c$ 
between $0.2$ to $1.2\,\vwind$, then approximately $95\%$ of the collisions exhibit perfect sticking and runaway growth of a 
boulder to a planetesimal is possible.  
Whether collisional fusion can occur in this range is a problem of material science under extreme conditions and
is the subject of ongoing research and a future paper. 

Recent studies \citep{gar+mer+gal+olc13,win+bir+orm+dul12} have pointed out 
that the PDF of collisional velocities is a crucial ingredient to the coagulation-fragmentation models.
In particular, \cite{win+bir+orm+dul12} have assumed the PDF of collisional velocities to be Maxwellian,
and have concluded that, by virtue of considering a PDF that is continuous at small values of its argument, 
growth by sticking is possible even if the sticking efficiency is determined by long-ranged intermolecular forces
(sticking with efficiency unity if the relative velocity of collisions is less than 5\,cm\,s$^{-1}$).
Here, we determine numerically the PDFs for the classes of collisions between boulders and dust grains and find that it 
cannot be simply described by a Maxwellian distribution - although it does have an exponential tail.
It is well known that in turbulent flows the PDF of the velocities of a tracer particle is Gaussian. 
We do not know of any study of the PDF of velocity of inertial particles
(particles that obey \eq{eq:dragp}) in turbulent flows, but it is
reasonable to assume that it would also be Gaussian.
If such an assumption holds, then we expect the PDF of collisional
velocities to have an exponential tail, so long as the size of particles
is not comparable. Were the colliding particles to be of roughly the same
size, the PDF may indeed have a power-law tail by virtue of intermittency.

In an earlier paper, \cite{sek+tak03} found that dust monomers advected by a steady laminar flow do not
collide with a spherical solid body with of radius much larger than the hydrodynamic length scale. 
The crucial limitation in their work was to assume the flow to be laminar. 
Here, we have considered turbulent flow and have obtained a different result, i.e., 
a significant percentage of the dust particles do hit the solid body with the PDF of collisional velocities 
peaking around the speed of the head wind. 

There exists an alternative scenario of planetesimal formation
\citep{joh+ois+mac+kla+hen+you07} in which the boulders are described
by the simple drag law \eq{eq:drag} but their back-reaction on the gas
is accounted for.
This is predicted to give rise to ``streaming instabilities'' which form boulder clusters around high pressure regions.  
Such clusters are then expected to coagulate by mutual gravitational interaction.  
In the light of the arguments presented in the present paper, this streaming instability scenario
requires further investigation.  This is because basic physical principles tell us that the description of the motion of
the boulder is inadequately described
by \eq{eq:drag}.
While the immersed boundary method can potentially solve this problem
we need to have massive computational resources to examine the fate of
many boulders.

We conclude by pointing out the limitations of our study. Firstly, here we confine ourselves to two dimensions.  
On the one hand, this has the virtue of permitting a larger range of $\Rep$ that can be easily accessed numerically.  
On the other hand we cannot capture the richness of particle fusion in the remaining dimension.  
However, we believe that this may imply that the growth of the particle we have studied to be a lower bound.  
Secondly, when collisional fusion starts operating the initial spherical object we study will not remain spherical. 
This may quantitatively affect further growth in a manner that depends on how the boulder tumbles through the disk. 
Thirdly, the turbulence in our flow is generated by external forcing.  It would be appropriate to use shearing-box simulations in three dimensions where the flow is driven by magneto-rotational instability.  We believe that these rather clear limitations do not detract from the robust results obtained in this study, which clarify the microphysical questions for a range of colliding materials and the computational fluid dynamics issues that will advance a sober assessment of planetesimal formation processes. 

\section*{Acknowledgments}
Financial support from the European Research Council under the AstroDyn Research Project 227952, and the Swedish 
Research Council grant  2011-5423 is gratefully acknowledged. 
J.S.W.\ thanks the Wenner-Gren and John Simon Guggenheim Foundations, and the Swedish Research Council.
We also thank the anonymous referee for his/her useful suggestions.

%\bibliography{turb_ref,sunref}
%\bibliographystyle{apj}
\end{document}